\documentclass[pra,twocolumn,showpacs,superscriptaddress,floatfix]{revtex4}
\usepackage{epsfig}
\usepackage{latexsym}
\usepackage{amsmath}
\usepackage{graphics}

\newcommand\br{\mathbf{r}}
\newcommand\bx{\mathbf{x}}

\newcommand\rd{\mathrm{d}}
\newcommand\te{\text{TE}}
\newcommand\tm{\text{TM}}

\begin{document}

\title{Retarded Casimir-Polder force on an atom near 
reflecting microstructures}
\author{Claudia Eberlein}
\author{Robert Zietal}
\affiliation{Dept of Physics \& Astronomy,
    University of Sussex,
     Falmer, Brighton BN1 9QH, England}
\date{\today}
\begin{abstract}
We derive the fully retarded energy shift of a neutral atom in two different
geometries useful for modelling etched microstructures. First we calculate
the energy shift due to a reflecting cylindrical wire, and then we work out
the energy shift due to a semi-infinite reflecting half-plane. We analyze
the results for the wire in various limits of the wire radius and the
distance of the atom from the wire, and obtain simple asymptotic expressions
useful for estimates. For the half-plane we find an exact representation of
the Casimir-Polder interaction in terms of a single, fast converging
integral, which is easy to evaluate numerically.
\end{abstract}


 \maketitle

\section{INTRODUCTION}
The explosive rate of developments in nanotechnology as well as in the
manipulation of cold atoms has meant that interest in atom-surface
interactions has increased strongly in recent years. What were once tiny,
elusive effects are now dominant interactions, or, as the case may be, a
major nuisance in some experimental set-ups. Motivated by a common type of
microstructure, which consists of a protruding ledge fabricated by
successive etching and possibly a thin electroplated top layer, we have
recently studied the force on a neutral atom in close proximity of
reflecting surfaces of either cylindrical geometry or that of a
semi-infinite half-plane \cite{nonret}. In the absence of free charges or
thermal excitations, the interaction of the atom with the microstructure is
dominated by Casimir-Polder forces \cite{casimir}, which are due to the
interaction of the atomic dipole with polarization fluctuations excited by
vacuum fluctuations of the electromagnetic field. If the atom is
sufficiently close to the surface of the microstructure, then the
interaction between the atomic dipole and the surface is purely
electrostatic and retardation can be neglected, which was the case
investigated in Ref.~\cite{nonret}. Then one does not need to quantize the
electromagnetic field, but can work with the classical Green's function of
Poisson's equation. The only difficulty lies then in the geometry of the
problem.

However, in experimental situations one more often finds that retardation is
in fact important, as the distance of the atom from the surface of the
microstructure is often commensurate or larger than the wavelength of a
typical atomic transition. This is the case we investigate here, again for
microstructures of two types of geometries: a cylindrical reflector of
radius $R$ and infinite length, and a reflecting half-plane.

Various versions of this problem have been studied before, both analytically
and numerically. Probably the first to consider the interaction between an
atom and a metallic wire, according to \cite{Barash}, was almost 75 years
ago Zel'dovich \cite{Zeldovich}. This problem was then revisited and
extended by Nabutovskii et al. \cite{Nabutovskii}, and subsequently by
Marvin et al. \cite{Marvin}. In Nabutovskii's paper a dielectric cylinder is
envisaged to be surrounded by a cylindrical shell of vacuum which in turn is
surrounded by a rarefied gas of polarizable particles. The interaction
energy of a single particle is then calculated through the work done by the
force (obtained from the stress tensor) due to the fluctuating
electromagnetic fields, in the limit of zero density of the surrounding
gas. The asymptotic results obtained there (Eq. (23) and Eq. (24) of
Ref.~\cite{Nabutovskii} ) are, according to Ref.~\cite{Barash}, valid only
for dilute dielectric materials; they diverge in the perfect-reflector
limit. 

On the other hand, the work by Marvin et al.\cite{Marvin}, motivated by
\cite{Mehl1, Mehl2} and based on a normal-mode expansion and a
linear-response formalism \cite{Langbein}, gives the same general
formula for the interaction between a point particle and a cylinder [their
Eq. (4.10)] as the equivalent result in \cite{Nabutovskii}.  We have no
reason to believe that the result in \cite{Marvin} is incorrect in the
perfect-conductor limit, as it reduces to our previous result \cite{nonret}
in the electrostatic limit. Moreover, Ref. \cite{Marvin} manages to recover
the original Casimir-Polder result \cite{casimir} in the large-radius limit
of the cylinder. This suggests that the general expression in
\cite{Nabutovskii} is probably correct, only that the perfect-conductor
limit does not commute with the asymptotic limit of the zero radius (or
large distance of the atom from the cylinder) studied there.  In the
small-radius limit, the result for the interaction between an atom and a
metallic filament, in both retarded and non-retarded limits, is also given
by \cite{Barash}.

Marvin et al.'s work \cite{Marvin} is certainly the most comprehensive, but
due to its generality it is also quite cumbersome to apply, which is mainly
done numerically for just a few examples \cite{Marvin2}. Further numerical
studies of the interaction of atoms with macroscopic cylinders can be found
in Refs.~\cite{Fussell,Boustimi1,Boustimi2,Blagov}.

By contrast, in this paper we are after a relatively simple theory that
allows one to estimate the force between an atom and a cylindrical reflector
at any distance and cylinder radius. To this end we are not interested in
the precise dependence of the interaction on material constants of the
reflector, and therefore we work with the model of a perfectly reflecting
surface. 

As discussed in Ref.~\cite{nonret}, we also determine the force between an
atom and a semi-infinite half-plane, in order to facilitate estimates for
common types of microstructures that consist of a ledge protruding from a
substrate. The Casimir-Polder interaction between an atom and such a
half-plane has also studied before, but only in the extreme retarded limit
of very large distances of the atom from the surface \cite{Brevik}. To the
best of our knowledge no formula for the interaction in the intermediate
region, when the distance of the atom from the surface is comparable to the
typical wavelength of an internal transition in the atom, has been derived
yet. Recent work of Mendes et al. \cite{Farina2}, dealing with wedges, does
not include the general result in the half-plane geometry as a limiting case
of a zero-angle wedge.




\section{FIELD QUANTIZATION AND THE ENERGY SHIFT}
The complete system of an atom interacting with the quantized
electromagnetic field is described by the Hamiltonian
\begin{equation}
H=H_{\rm{Atom}}+H_{\rm{Field}}+H_{\rm{Int}}\;.
\end{equation}
We choose to work with $\pmb{\mu}\cdot{\bf E}$ coupling, i.e. our interaction
Hamiltonian is
\begin{equation}
H_{\rm{Int}}=-\pmb{\mu}\cdot{\bf E}\;.
\label{Hint}
\end{equation}
Quantization of the electromagnetic field is done by way of a normal-mode
expansion of the vector potential in terms of photon annihilation and
creation operators for each mode $\lambda$ and polarization $\sigma$,
\begin{equation}
\mathbf{A}(\br,t)=\sum_{\lambda,\sigma}\frac{1}{\sqrt{2\varepsilon_0
\omega_\lambda}}
\left[a_\lambda^{(\sigma)}{\bf F}_{\lambda}^{(\sigma)}(\br)\;
e^{-{\rm i}\omega t}  + \mbox{h.c.}\right].
\end{equation}
To describe a mode we use the composite index $\lambda$ instead of a wave
vector, as we shall be working in cylindrical coordinates where the quantum
number of the azimuthal part of the mode function is discrete, but the other
two are continuous.  We work in Coulomb gauge, $\pmb{\nabla}\cdot{\bf
A}(\br)$, so that the normal modes ${\bf F}(\br)$ satisfy the Helmholtz
equation,
\begin{equation}
(\pmb{\nabla}^2+\omega^2)\,\mathbf{F}(\br)=0\;.
\label{helm}
\end{equation}

The energy level shift due to the interaction (\ref{Hint}) can be calculated
perturbatively. For our system in state $|i;0\rangle$, i.e. the atom in
state $|i\rangle$ and the electromagnetic field in its vacuum state
$|0\rangle$, the lowest non-vanishing order of perturbation theory is the
second, so that
\begin{equation}
\Delta W = \sum_{j \neq i} \frac{\left|\left\langle j;1_\lambda^{(\sigma)}
\left|-\pmb{\mu}\cdot{\bf E}\right| i;0 \right\rangle
\right|^2}{E_i - (E_j+\omega_\lambda)}\;.
\end{equation}
As the relevant field modes can be expected to vary slowly over the size of
the atom, we make the electric dipole approximation, which simplifies the
expression for the energy shift to
\begin{equation}
\Delta W = -\sum_{\lambda,\sigma,j \neq i} 
\frac{\omega_\lambda}{2\varepsilon_0} \; \frac{\left|
\langle j|\pmb{\mu}|i\rangle \cdot {\bf F}^{(\sigma)*}_\lambda(\br)
\right|^2}{E_{ji}+\omega_\lambda}\;,
\label{dW}
\end{equation}
where we have introduced the abbreviation $E_{ji}\equiv E_j - E_i$. For
brevity and presentational clarity we shall henceforth also abbreviate the
matrix elements of the atomic dipole moment as
\begin{equation}
\left|\pmb{\mu}\right|\equiv\left|\left\langle j\left| \pmb{\mu} 
\right| i \right\rangle\right|\;.
\label{mu}
\end{equation}

\section{ENERGY SHIFT NEAR A PERFECTLY REFLECTING WIRE}
\label{sec3}
\begin{figure}[th]
\vspace*{3mm}
\begin{center}
\centerline{\epsfig{figure=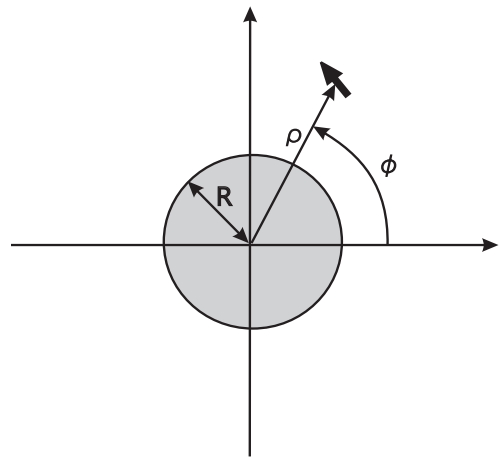,
width= 6.5cm, height=6.0 cm}}
\end{center}
\vspace*{-5mm} \caption{Atomic electric dipole moment in the vicinity of a
perfectly reflecting cylinder of radius $R$. The normal modes
$\mathbf{F}_\lambda^{(\sigma)}(\bx)$ in this geometry are given by
Eqs. (\ref{CylinderTE}) and (\ref{CylinderTM}). }
\label{fig:1}
\end{figure}
First we wish to calculate the energy shift of an atom near a perfectly
reflecting and infinitely long cylindrical wire of radius $R$. It is
advantageous to work in cylindrical coordinates, cf. Fig.~\ref{fig:1}.

In order to find two independent transverse vector field solutions of
Eq.~(\ref{helm}), we make use of the representation theorem for the vector
Helmholtz equation \cite[10.411]{GR}. If $\Phi(\bx)$ is a solution of the
scalar Helmholtz equation then the two independent solutions of the vector
equation are given by
\begin{eqnarray}
\mathbf{F}^{(1)}(\br) &\!\!=&\!\! \left(\mathbf{\nabla}\times 
\hat{\mathbf{e}}_z\right)\Phi(\br)\;,\label{sol1} \\
\mathbf{F}^{(2)}(\br) &\!\!=&\!\! \frac{1}{\omega}\mathbf{\nabla}
\times\left(\mathbf{\nabla}\times \hat{\mathbf{e}}_z\right)\Phi(\br)\;.
\label{sol2}
\end{eqnarray}
The particular choice of the constant unit vector $\hat{\mathbf{e}}_z$ is
motivated by the symmetry of our problem and lets us to identify the
solutions $\mathbf{F}^{(1)}(\br)$ and $\mathbf{F}^{(2)}(\br)$ with the
transverse electric (TE) and transverse magnetic (TM) modes, respectively.
In cylindrical coordinates the scalar Helmholtz equation
has the solutions of the form
\begin{equation}
\Phi(\rho,\phi,z)=N\left[ \;\cos\delta_m J_m(k\rho) + \sin\delta_m
  Y_m(k\rho) \; \right]e^{{\rm i}m\phi+{\rm i}\kappa z}
\label{scalarnorm}
\end{equation}
where $J_m(k\rho)$ and $Y_m(k\rho)$ are Bessel functions of the first and
second kind \cite[{\S}9]{AS}. The separation constants satisfy
$\omega^2=k^2+\kappa^2$, and $m$ is an integer. The phase shifts $\delta_m$
describe the superposition of regular and irregular solutions. In free space
only regular solutions $J_m(k\rho)$ are admissible, and $\delta_m=0$. In the
presence of the perfectly reflecting wire, the phase shifts serve to make the
electromagnetic fields satisfy the boundary conditions on the surface of the
wire.  The normalization constant $N$ is chosen such that
\begin{equation}
\int \rd^3\br\; \mathbf{F}^{(\sigma)*}_{\lambda'}(\br)\cdot
\mathbf{F}^{(\sigma)}_{\lambda}(\br)
=\delta_{mm'}\delta(\kappa-\kappa')\frac{\delta{(k-k')}}{\sqrt{kk'}}
\label{vectornorm}
\end{equation}
is met. Setting $\cos\delta_m=1$, $\sin\delta_m=0$, one can derive quite
easily that $N=(2\pi k)^{-1}$.

On the surface of a perfect conductor, the tangential components of the
electric field and the normal component of the magnetic field
vanish. Therefore, at the surface $\rho=R$ of the cylindrical wire we must
have $E_\phi=0=E_z$ and $B_\rho=0.$ These boundary conditions determine the
phase shifts as
\begin{eqnarray}
\tan\delta_m^{\te} = -\frac{J'_m(kR)}{Y'_m(kR)},\;\tan\delta_m^{\tm} 
= -\frac{J_m(kR)}{Y_m(kR)}\;. \label{PhaseShifts}
\end{eqnarray}
According to Eqs.~(\ref{sol1}), (\ref{sol2}), and (\ref{scalarnorm}), the
normalized mode functions $\mathbf{F}_\lambda^{(\sigma)}(\br)$,
$\lambda=\{k,m,\kappa\}$, that satisfy the boundary conditions at $\rho=R$
are given by
\begin{eqnarray}
&&\mathbf{F}_\lambda^{\te}(\rho,\phi,z) \nonumber\\
&&=\frac{1}{2\pi}
\left[\frac{{\rm i}m}{k\rho}\frac{J_m(k\rho)Y'_m(kR)-Y_m(k\rho)J'_m(kR)}{
\sqrt{J_m^{'2}(kR)+Y_m^{'2}(kR)}}\;\hat{\mathbf{e}}_\rho\right.
\label{CylinderTE}\\
&&\ \ \ - \left.\frac{J'_m(k\rho)Y'_m(kR)-Y'_m(k\rho)J'_m(kR)}{
\sqrt{J^{'2}_m(kR)+Y^{'2}_m(kR)}}\;\hat{\mathbf{e}}_\phi
\right]e^{{\rm i}m\phi+{\rm i}\kappa z},
\nonumber
\end{eqnarray}
\begin{eqnarray}
&&\mathbf{F}_\lambda^{\tm}(\rho,\phi,z) \nonumber\\&&=\frac{1}{2\pi} \left[
\frac{i\kappa}{\omega} \frac{J'_m(k\rho)Y_m(kR)-
Y'_m(k\rho)J_m(kR)}{\sqrt{J_m^2(kR)+Y_m^2(kR)}}\;
\hat{\mathbf{e}}_\rho\right.\nonumber\\
&&\ \ \ -\frac{m\kappa}{\omega k \rho} \frac{J_m(k\rho)Y_m(kR)-
Y_m(k\rho)J_m(kR)}{\sqrt{J_m^2(kR)+Y_m^2(kR)}}\;\hat{\mathbf{e}}_\phi
\label{CylinderTM}\\
&&\ \ \ +\left.\frac{k}{\omega} \frac{J_m(k\rho)Y_m(kR)-
Y_m(k\rho)J_m(kR)}{\sqrt{J_m^2(kR)+Y_m^2(kR}}\;\hat{\mathbf{e}}_z
\right]e^{{\rm i}m\phi+{\rm i}\kappa z}.
\nonumber
\end{eqnarray}
These mode functions can now be substituted into Eq.~(\ref{dW}) for
obtaining the energy shift of an atom positioned at $\br=(\rho,\phi,z)$.
However, what we want to calculate here is only the correction to the energy
shift caused by the presence of a perfectly conducting surface, rather than
the whole energy shift due to the coupling of the atom to the fluctuating
vacuum field, which would include the free-space Lamb shift. Therefore we
need to subtract the energy shift caused by the vacuum fluctuations of the
electromagnetic field in free space, which is obtained by either letting the
phase shifts $\delta_m\rightarrow 0$ or equivalently taking the limit
$R\rightarrow 0$. In the limit of vanishing radius $R$ of the cylinder the
behaviour of the mode functions (\ref{CylinderTE}), (\ref{CylinderTM}) is
dominated by the singular behaviour of $Y_m(kR)$ and $Y'_m(kR)$ at the
origin, which causes the phase shifts (\ref{PhaseShifts}) to
vanish. The renormalized energy
shift $\Delta W^{\rm{ren}}=\Delta W -\lim_{R\rightarrow 0}\Delta W$ 
is found to be of the form 
\begin{equation} 
\Delta W^{\rm{ren}} = - \frac{1}{4\pi\varepsilon_0} \sum_{j \neq i} \left(
\Xi_\rho |\mu_\rho|^2 + \Xi_\phi |\mu_\phi|^2 + \Xi_z |\mu_z|^2 \right)
\label{generalShift}
\end{equation}
with
\begin{widetext}
\begin{eqnarray}
\Xi_\rho &\!\!=&\!\!\frac2\pi \sum_{m=0}^\infty\!\! \text{ }' 
\int_0^\infty \rd k\; k\int_0^\infty \rd\kappa 
\;\frac{\omega}{E_{ji}+\omega}
\left\{
\left(\frac{m}{k\rho}\right)^2\left[\frac{\big(J_m(k\rho)Y'_m(kR)
- Y_m(k\rho)J'_m(kR)\big)^2}{J^{'2}_m(kR)+Y^{'2}_m(kR)}-J^2_m(k\rho)\right]
\right.\nonumber\\
&\!\!&\hspace*{35mm}+\left.
\left(\frac{\kappa}{\omega}\right)^2\left[\frac{\big(J'_m(k\rho)Y_m(kR)
-Y'_m(k\rho)J_m(kR)\big)^2}{J_m^2(kR)+Y_m^2(kR)}-J'^2_m(k\rho)\right]
\right\},
\label{FreeKsiRho}
\\
\Xi_\phi &\!\!=&\!\!\frac2\pi \sum_{m=0}^\infty\!\! \text{ }' 
\int_0^\infty \rd k\; k\int_0^\infty \rd\kappa 
\;\frac{\omega}{E_{ji}+\omega}
\left\{
\left[\frac{\big(J'_m(k\rho)Y'_m(kR) - Y'_m(k\rho)J'_m(kR)\big)^2}{
J^{'2}_m(kR)+Y^{'2}_m(kR)}-J'^2_m(k\rho)\right]
\right.\nonumber\\
&\!\!&\hspace*{35mm}+\left.\left(\frac{m}{k\rho}\frac{\kappa}{\omega}\right)^2
\left[\frac{\big(J_m(k\rho)Y_m(kR)- Y_m(k\rho)J_m(kR)\big)^2}{
J_m^2(kR)+Y_m^2(kR)}-J^2_m(k\rho)\right]
\right\},
\label{FreeKsiPhi}
\\
\Xi_z &\!\!=&\!\!\frac2\pi\sum_{m=0}^\infty\!\! \text{ }' 
\int_0^\infty \rd k\; k\int_0^\infty \rd\kappa 
\;\frac{\omega}{E_{ji}+\omega}
\left\{
\left(\frac{k}{\omega}\right)^2\left[\frac{\big(J_m(k\rho)Y_m(kR)
-Y_m(k\rho)J_m(kR)\big)^2}{J_m^2(kR)+Y_m^2(kR)}-J^2_m(k\rho)\right]
\right\}
\label{FreeKsiZ}
\end{eqnarray}
\end{widetext}
where the primes on the sums indicate that the ${m=0}$ term is weighted by
an additional factor of 1/2. It appears that the $\kappa$ integrals fail to
converge, but this is a common feature in such calculations caused by the
dipole approximation, see e.g. \cite{casimir}. As we shall see, convergence
is in fact brought about by the Bessel functions, which come to bear if the
$k$ integral is replaced by an integral over $\omega=\sqrt{\kappa^2+k^2}$. 

As the Bessel functions $J_m(x)$ and $Y_m(x)$ are both oscillatory for large
$x$, we wish to rotate the integration contour in the complex $k$ plane, in
order to get an integrand that is exponentially damped for large
arguments. To this end we introduce the Hankel functions
$H_m^{(1)}(x)=J_m(x)+iY_m(x)$ and
$H_m^{(2)}(x)=[H_m^{(1)}(x)]^*=J_m(x)-iY_m(x)$, in terms of which we can
rewrite the energy level shift in such a form that there are no poles in the
first quadrant of the complex $k$ plane, as is required for the rotation of
the  integration contour. This step greatly simplifies further analysis.
\begin{widetext}
\begin{eqnarray}
\Xi _\rho &\!\!=&\!\! -\mathrm{Re}\ \frac2\pi \sum_{m=0}^\infty\!\! \text{ }' 
\int_0^\infty \rd k\; k\int_0^\infty \rd\kappa 
\;\frac{\omega}{E_{ji}+\omega}
\left\{
 \frac{\kappa^2}{\omega^2} [H_m^{\prime(1)}(
k\rho)]^2 \frac{J_m(kR)}{H^{(1)}_m(kR)}+
\frac{m^2}{k^2\rho^2}[H^{(1)}_m(k\rho)]^2 
\frac{J'_m(kR)}{H_m^{\prime (1)}(kR)}
\right\},
\label{KsiRho}
\\
\Xi _{\phi } &\!\!=&\!\!-\mathrm{Re}\ \frac2\pi \sum_{m=0}^\infty\!\!\text{ }' 
\int_0^\infty \rd k\;k\int_0^\infty \rd\kappa 
\;\frac{\omega}{E_{ji}+\omega}
\left\{ 
[H_m^{\prime(1)}(
k\rho)]^2 \frac{J'_m(kR)}{H^{\prime (1)}_m(kR)}+
\frac{m^2}{k^2\rho^2}\frac{\kappa^2}{\omega^2}[H^{(1)}_m(k\rho)]^2 
\frac{J_m(kR)}{H_m^{(1)}(kR)}\right\},
\label{KsiPhi}
\\
\Xi _{z} &\!\!=&\!\!-\mathrm{Re}\ \frac2\pi  \sum_{m=0}^\infty\!\! \text{ }' 
\int_0^\infty \rd k\;k\int_0^\infty \rd\kappa 
\;\frac{\omega}{E_{ji}+\omega}
\left\{
\frac{k^2}{\omega^2} [H_m^{(1)}(
k\rho)]^2 \frac{J_m(kR)}{H^{(1)}_m(kR)}
\right\}.
\label{KsiZ}
\end{eqnarray}
\end{widetext}
We now transform the $k$ integration in Eqs. (\ref{KsiRho})--(\ref{KsiZ})
into an integration over $\omega=\sqrt{\kappa^2+k^2}$, and note that on the
interval $0\leq\omega\leq\kappa$ the integrands become pure imaginary and
therefore do not contribute if added to the real part of the integral. We can
therefore shift the lower limit down to the origin
\begin{equation}
\int_\kappa^\infty \rd \omega\longrightarrow\int_0^\infty \rd\omega\;
\end{equation}
without affecting the result. Further, we note that the functions
$H_m^{(1)}(z)$ and $H_m^{\prime(1)}(z)$ have no zeros in the first quadrant
of the complex plane \cite[Fig. 9.6]{AS}, so that the contour of the
$\omega$-integration can be rotated from the positive real to the positive
imaginary axis, $\omega\rightarrow{\rm i}\omega$. Then the oscillatory
Bessel functions turn into the modified Bessel functions according to
\cite[9.6.3 \& 5]{AS}
\begin{eqnarray}
J_m({\rm i}z)&\!\!=&\!\! e^{{\rm i}m\pi /2}I_m(z), \\
H_m^{(1)}({\rm i}z) &\!\!=&\!\! -\frac{2{\rm i}}{\pi} 
e^{-{\rm i}m\pi /2}K_m(z)\;.
\end{eqnarray}
Taking the real part and going to polar coordinates, where the angle
integrals are elementary, we find that
\begin{widetext}
\begin{eqnarray}
\Xi _\rho &\!\!=&\!\! \frac2\pi\sum_{m=0}^\infty\!\! \text{ }' 
\int_0^\infty \rd k\;k
\left\{
 \left(\sqrt{E_{ji}^2+k^2}-E_{ji}\right)  
\frac{I_m(kR)}{K_m(kR)}[K'_m(k\rho)]^2+
\frac{m^2}{k^2\rho^2}\left(\frac{E_{ji}^2}{\sqrt{E_{ji}^2+k^2}}-E_{ji}\right) 
\frac{I'_m(kR)}{K'_m(kR)}[K_m(k\rho)]^2
\right\},\nonumber\\
\label{FinalC1}
\\
\Xi _\phi &\!\!=&\!\! \frac2\pi\sum_{m=0}^\infty\!\! \text{ }' 
\int_0^\infty \rd k\;k
\left\{
 \left(\frac{E_{ji}^2}{\sqrt{E_{ji}^2+k^2}}-E_{ji}\right) 
\frac{I'_m(kR)}{K'_m(kR)}[K'_m(k\rho)]^2
 +\frac{m^2}{k^2\rho^2}\left(\sqrt{E_{ji}^2+k^2}-E_{ji}\right) 
\frac{I_m(kR)}{K_m(kR)}[K_m(k\rho)]^2
\right\},\nonumber\\
\label{FinalC2}\\
\Xi _{z} &\!\!=&\!\!\frac2\pi\sum_{m=0}^\infty\!\! \text{ }' 
\int_0^\infty \rd k\;k
\left\{
\frac{k^2}{\sqrt{E_{ji}^2+k^2}}\frac{I_m(kR)}{K_m(kR)}[K_m(k\rho)]^2 
\right\}.
\nonumber\\\label{FinalC3}
\end{eqnarray}
\end{widetext}
Note that the effect of our manipulations has been that the integration
variable $k$ in Eqs.~(\ref{FinalC1})--(\ref{FinalC3}) has been rotated by
$\pi/2$ in the complex plane compared to Eqs.~(\ref{KsiRho})--(\ref{KsiZ}).

The final result for the energy shift, Eq.~(\ref{generalShift}) with
Eqs.~(\ref{FinalC1})--(\ref{FinalC3}), is a sum over a series of rapidly
converging integrals, which, unlike
Eqs.~(\ref{FreeKsiRho})--(\ref{FreeKsiZ}), is reasonably easily evaluated
numerically. However, as the functions $\Xi_{\rho,\phi,z}(E_{ji}, d, R)$
are quite cumbersome and it is not possible to find exact closed form
expressions for them, we now look at their asymptotics in various limiting
cases, which is very useful for analytical estimates. 

\subsection{Asymptotic regimes}
There are three length scales in the problem: the distance of the atom from
the surface of the cylinder $d=\rho-R$, the radius of the cylindrical wire
$R$, and the wavelength of a typical transition in the atom $\lambda_{ji}
\propto 1/E_{ji}$. Accordingly we get six different asymptotic regimes,
three non-retarded and three retarded. The criterion as to whether
retardation matters is the relative size of the distance $d$ of the atom
from the surface and the wavelength $\lambda_{ji}$ of a typical transition:
if the atom is very close to the surface then its interaction with the
surface is entirely electrostatic \cite{nonret}, whereas retardation begins
to play a role once $d\sim \lambda_{ji}$ or larger, because then the
internal state of the atom is then subject to non-negligible evolution
during the time a virtual photon mediating the interaction would take to
travel from the atom to the surface and back. First we shall deal with the
three non-retarded cases, and then with the three retarded ones.

\subsubsection{$d\ll R \ll \lambda_{ji}$}\label{case1}
If $\lambda_{ji}$ is larger than any other lengthscale, we can take the
limit $E_{ji}\rightarrow 0$ in Eqs.~(\ref{FinalC1})--(\ref{FinalC3}). This
gives the same result as a purely electrostatic calculation \cite{nonret}.
If the distance $d$ of the atom from the surface is small, then the atom
does not feel the curvature of the surface, and one expects to get the same
energy shift as one would close to a plane surface. This is indeed the
result we get when we take the limit $d\rightarrow 0$ by using uniform
asymptotic expansions for the Bessel functions \cite{nonret}; we obtain
\begin{eqnarray}
\Xi _\rho \approx \frac{1}{8d^3}\;,\;\;\Xi _\phi \approx 
\frac{1}{16d^3}\;,\;\;\Xi _z  \approx \frac{1}{16d^3}\;. \label{plane}
\end{eqnarray}

\subsubsection{$d \ll \lambda_{ji} \ll R$}\label{case5}
In this regime the energy shift behaves in exactly the same way as in the
previous case, because the radius of the wire has no influence on
retardation, so that the relative size of $R$ and $\lambda_{ji}$ does not
matter. All that matters is that the distance $d$ of the atom from the
cylinder is still much less than the wavelength $\lambda_{ji}$ of the
relevant transition in the atom. In mathematical terms, the electrostatic
limit ($E_{ji}\rightarrow 0$) and the large-radius limit
($R\rightarrow\infty$) of the energy shift commute.

The limit of large radius was studied in great detail in
\cite{Marvin}. Application of the summation formula derived in Appendix A of
\cite{Marvin} to Eqs.~(\ref{FinalC1})-(\ref{FinalC3}) leads to the original
Casimir-Polder result \cite{casimir} for the interaction between an atom and a
plane, perfectly reflecting mirror:
\begin{eqnarray}
\Xi_\rho &\!\!=&\!\! \frac{1}{2\pi d^3}\int_0^\infty \rd\eta 
\frac{e^{-2d E_{ji}\eta}}{(1+\eta^2)^2}\;, \label{CPperp} \\
\Xi_\phi=\Xi_z &\!\!=&\!\! \frac{1}{2\pi d^3}\int_0^\infty \rd\eta 
\frac{e^{-2d E_{ji}\eta}}{(1+\eta^2)^2}\frac{1-\eta^2}{1+\eta^2}\;.
\label{CPpara}
\end{eqnarray}
If we now take $\lambda_{ji}$ to be much greater than $d$, we reproduce the
result (\ref{plane}) of the previous section.

\subsubsection{$R\ll d \ll \lambda_{ji}$}\label{case2}
In this case we again start by taking the limit $E_{ji}\rightarrow 0$ in
Eqs.~(\ref{FinalC1})--(\ref{FinalC3}) and obtain the electrostatic
expression derived in \cite{nonret}. In the limit of the radius of the wire
being much smaller than the distance $d$, the energy shift is dominated by
summand with lowest $m$ in Eqs.~(\ref{FinalC1})--(\ref{FinalC3})
\cite{nonret}. Asymptotically one gets
\begin{eqnarray}
\Xi _\rho  \propto  \frac{1}{d^3\ln d},\;\; \Xi _\phi &\!\! \propto &\!\! 
\frac{R^2}{d^5},\;\;\Xi _z \propto \frac{1}{d^3\ln d}\;, \nonumber
\end{eqnarray}
which is not very helpful numerically though, as logarithmic series converge
only very slowly.

\subsubsection{$\lambda_{ji} \ll d \ll R$}\label{case3}
\begin{figure}[t]
\vspace*{-10mm}
\begin{center}
\centerline{\epsfig{figure=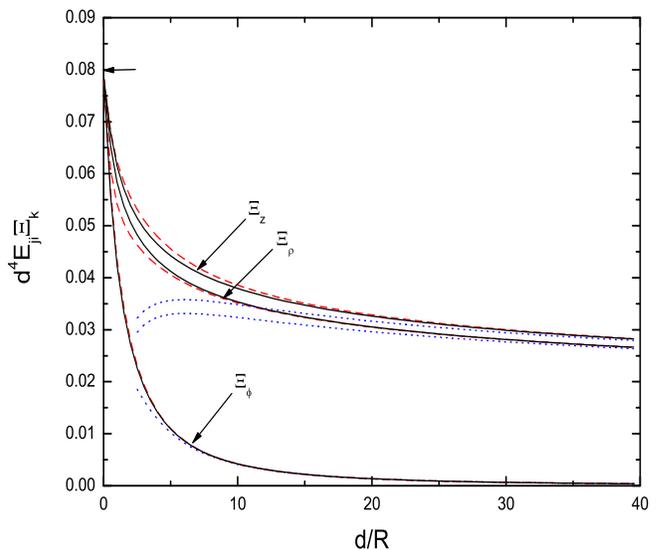,
width= 10 cm, height= 9.0 cm}}
\end{center}
\vspace*{-15mm} \caption{The contributions to the energy shift in the
  retarded limit due
to the three components of the atomic dipole, multiplied by
$E_{ji}d^4$. Solid lines represent the results of exact numerical
integration of Eqs.~(\ref{FinalC1})--(\ref{FinalC3}) in the limit
$E_{ji}\rightarrow\infty$, whereas the dashed (red) lines represent the
approximations (\ref{A})--(\ref{C}). For large $d$ the asymptotic behaviour
is dominated by the lowest $m$ terms in the sums, given by
(\ref{A2})--(\ref{C2}) and shown as dotted (blue) lines.  The arrow on the
vertical axis indicates the exact value in the limit $d\rightarrow 0$,
Eq. (\ref{PlaneCP}).}
\label{fig:2}
\end{figure}

When $\lambda_{ji}$ is smaller than the distance $d$ of the atom to the
surface of the wire, then the interaction is manifestly retarded. As
$\lambda_{ji}$ is the smallest of the three lengthscales, we first take the
limit $\lambda_{ji}\rightarrow 0$, i.e.~$E_{ji}\rightarrow\infty$, in
Eqs.~(\ref{FinalC1})--(\ref{FinalC3}) and find that the leading terms in all
three integrals go as $1/E_{ji}$. The remaining integration over $k$ is then
quite similar to those found in the non-relativistic calculation in
\cite{nonret} and can be tackled by the same means. Scaling $k$ to
$x=k\rho/m$ and realizing that the dominant contributions to the integrals
and sums come from large $x$ and large $m$, one can approximate the Bessel
Functions by their uniform asymptotic expansions and then gets a geometric
series, which is easy to sum. In this way one finds the following
approximations
\begin{eqnarray}
\Xi _\rho &\!\! \approx &\!\! \frac{1}{2\pi E_{ji}\rho^4}\left\{ 
\rho^4\int_0^\infty \rd k\, k^3
 \frac{I_0(kR)}{K_0(kR)}[K_1(k\rho)]^2\right.
 \label{A}\\ &\!\! + &\!\!
\left.\int_0^\infty \rd x\;x\left(\sqrt{1+x^2}+\frac{1}{\sqrt{1+x^2}}\right)
\frac{A(A^2+4A+1)}{(A-1)^4}\right\},\nonumber\\
\Xi _\phi &\!\! \approx &\!\! \frac{1}{2\pi E_{ji}\rho^4}\left\{
\rho^4\int_0^\infty \rd k\, k^3
 \frac{I_1(kR)}{K_1(kR)}[K_1(k\rho)]^2\right.
\label{B}\\ &\!\! + &\!\!
\left.\int_0^\infty \rd x\;x\left(\sqrt{1+x^2}+\frac{1}{\sqrt{1+x^2}}\right)
\frac{A(A^2+4A+1)}{(A-1)^4}\right\},\nonumber\\
\Xi _z &\!\! \approx &\!\! \frac{1}{\pi E_{ji}\rho^4}\left\{
\rho^4\int_0^\infty \rd k\, k^3
 \frac{I_0(kR)}{K_0(kR)}[K_0(k\rho)]^2\right.
\nonumber\\ &\!\! + &\!\!
\left.\int_0^\infty \rd x\;\frac{x^3}{\sqrt{1+x^2}}\,
\frac{A(A^2+4A+1)}{(A-1)^4}\right\},\label{C}
\end{eqnarray}
with $A(x)$ given by
\begin{eqnarray}
A(x) &\!\! = &\!\! \left( \frac{R}{\rho} \right)^2 \left( 
\frac{1+\sqrt{1+x^2}}{1+\sqrt{1+x^2\frac{R^2}{\rho^2}}} \right)^2 
\nonumber\\ &\!\! \times&\!\!\exp \left[ 2\left( 
\sqrt{1+x^2\frac{R^2}{\rho^2}} - \sqrt{1+x^2} \right) \right].
\end{eqnarray}
These are easy to evaluate numerically and provide a reasonable
approximation to the energy shift in the retarded limit, as shown in
Fig. \ref{fig:2}.
In the limit of the distance $d=\rho-R$ being much smaller than the radius
$R$ of the wire, the above approximations yield
\begin{equation}
\Xi _\rho \approx \Xi _\phi \approx \Xi _z  
\approx \frac{1}{4\pi d^4}\frac{1}{E_{ji}} \;,\label{PlaneCP}
\end{equation}
which agrees with the retarded energy shift of an atom in front of a
perfectly reflecting plane mirror, as calculated by Casimir and Polder
\cite{casimir}. This is what one would expect because an atom that is very
close to the surface is not susceptible to the curvature of the surface.

\subsubsection{$\lambda_{ji} \ll R \ll d$}\label{case4}
In this case we again start by taking the limit $E_{ji}\rightarrow\infty$ in
Eqs.~(\ref{FinalC1})--(\ref{FinalC3}), which gives a leading order
contribution proportional to $1/E_{ji}$. For distances $d$ much larger than
the wire radius $R$ the dominant contribution to the sum then comes from the
summands with the lowest $m$, so that we need consider only those,
\begin{eqnarray}
&&\hspace*{-10mm}\Xi _\rho \approx \frac{1}{2\pi E_{ji}}\left\{ 
\int_0^\infty \rd k\, k^3
 \frac{I_0(kR)}{K_0(kR)}[K_1(k\rho)]^2\right.
 \label{A2}\\ && \hspace*{15mm}
\left.-2\int_0^\infty \rd k\, \frac{k}{\rho^2}
 \frac{I_1'(kR)}{K_1'(kR)}[K_1(k\rho)]^2
\right\},\nonumber\\
&&\hspace*{-10mm}\Xi _\phi \approx \frac{1}{2\pi E_{ji}}\left\{
\int_0^\infty \rd k\,k\! \left(k^2+\frac{2}{\rho^2}\right)\!
 \frac{I_1(kR)}{K_1(kR)}[K_1(k\rho)]^2\right.
\label{B2}\\ 
&& \hspace*{15mm}
\left.-2\int_0^\infty \rd k\, k^3
 \frac{I_1'(kR)}{K_1'(kR)}[K_1'(k\rho)]^2
\right\},\nonumber\\
&&\hspace*{-10mm}\Xi _z \approx \frac{1}{\pi E_{ji}}
\int_0^\infty \rd k\, k^3
 \frac{I_0(kR)}{K_0(kR)}[K_0(k\rho)]^2\;.
\label{C2}
\end{eqnarray}
The dotted lines in Fig. \ref{fig:2} show that these are indeed good
approximations for large $d/R$.  Their leading-order behaviour is
\begin{eqnarray}
\Xi _\rho  \propto  \frac{1}{E_{ji}}\frac{1}{d^4\ln d},\;\; 
\Xi _\phi &\!\! \propto &\!\! \frac{1}{E_{ji}}\frac{R^2}{d^6},\;\;
\Xi _z \propto \frac{1}{E_{ji}}\frac{1}{d^4\ln d}\;, \nonumber
\end{eqnarray}
which is in full agreement with the asymptotic results by \cite{Barash},
even though those are for a metallic wire characterized by a plasma
frequency. This is because in the retarded limit the interaction between the
atom and the surface depends, to leading order, only on the static
polarizability.

As in the electrostatic case, the contributions due to the $\rho$ and $z$
components of the atomic dipole fall off less rapidly than the $\phi$
contribution. We also note that, just as in the non-retarded case, the
series in powers of $1/\ln d$ converge too slowly to be of any practical
use, so that estimates must be made with Eqs.~(\ref{A2})--(\ref{C2}).
 
\subsubsection{$R \ll \lambda_{ji} \ll d$}\label{case6}
As in the non-retarded cases, the limit of vanishing radius ($R\rightarrow
0$) and the retarded limit ($E_{ji}\rightarrow \infty$) commute, and we
recover the results of the previous section,
Eqs.~(\ref{A2})--(\ref{C2}). This is another manifestation of the fact that
the criterion of whether the interaction is retarded depends solely on the
distance $d$ between an atom and the surface of the wire, and that the
relative size of geometrical features and the wavelength $\lambda_{ji}$ is
irrelevant. This means in particular that there are no resonance effects for
$\lambda_{ji}$ co\"{\i}nciding with the wire radius $R$.

\subsection{Numerical results}
\begin{figure}
\vspace*{-6mm}
\begin{center}
\centerline{\epsfig{figure=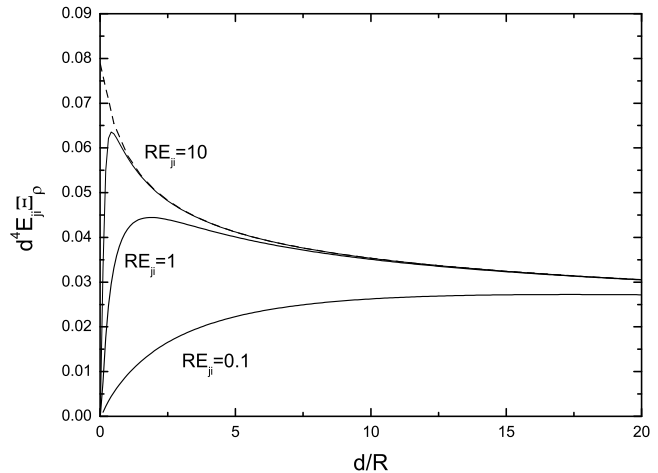,
width= 10cm}}
\end{center}
\vspace*{-10mm} \caption{The contribution (\ref{FinalC1}) to the energy
  shift (\ref{generalShift}) due to the $\rho$ component of the dipole for
  various typical transition frequencies $E_{ji}$. The dashed line is this
  contribution in the retarded limit $E_{ji}\rightarrow\infty$.}
\label{fig:rho}
\end{figure}

\begin{figure}
\vspace*{-6mm}
\begin{center}
\centerline{\epsfig{figure=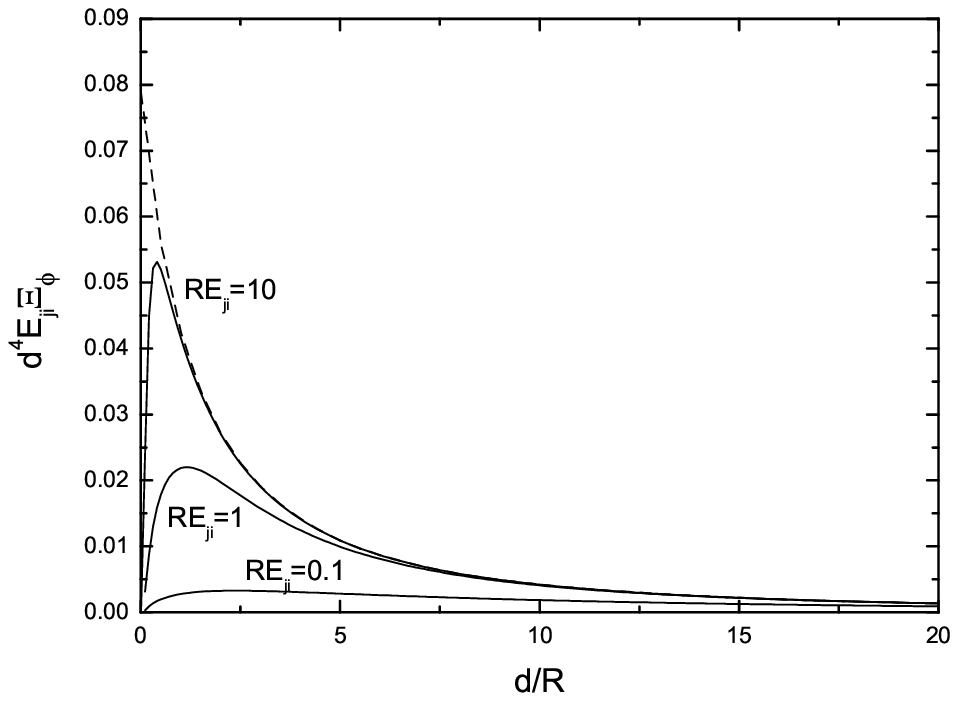,
width= 10cm}}
\end{center}
\vspace*{-10mm} \caption{The contribution (\ref{FinalC2}) to the energy
  shift (\ref{generalShift}) due to the $\phi$ component of the dipole for
  various typical transition frequencies $E_{ji}$. The dashed line is this
  contribution in the retarded limit $E_{ji}\rightarrow\infty$.}
\label{fig:phi}
\end{figure}

\begin{figure}
\vspace*{-6mm}
\begin{center}
\centerline{\epsfig{figure=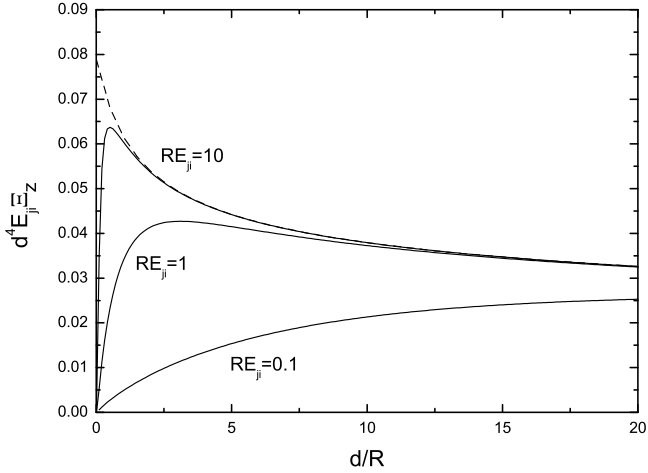,
width= 10cm}}
\end{center}
\vspace*{-10mm} \caption{The contribution (\ref{FinalC3}) to the energy
  shift (\ref{generalShift}) due to the $z$ component of the dipole for
  various typical transition frequencies $E_{ji}$. The dashed line is this
  contribution in the retarded limit $E_{ji}\rightarrow\infty$.}
\label{fig:zet}
\end{figure}

For intermediate parameter ranges one has to evaluate
Eqs.~(\ref{FinalC1})--(\ref{FinalC3}) numerically. This is straightforward,
and one can employ standard software packages like Mathematica or Maple. The
numerical convergence of Eqs.~(\ref{FinalC1})--(\ref{FinalC3}) is very good,
although more terms are needed for small distances $d$ than for large
distances.  Figs.~\ref{fig:rho}--\ref{fig:zet} show the contributions by the
$\rho$, $\phi$, and $z$ components of the atomic dipole to the energy shift
(\ref{generalShift}) for various values of the typical transition frequency
$E_{ji}$ in the atom. We give the distance $d$ and the transition wavelength
$1/E_{ji}$ in units of the wire radius $R$. For plotting we have factored
out of $\Xi_{\rho,\phi,z}$ the asymptotic functional dependence of the shift
in front of a plane mirror, Eq.~(\ref{PlaneCP}).

\begin{figure}
\vspace*{-6mm}
\begin{center}
\centerline{\epsfig{figure=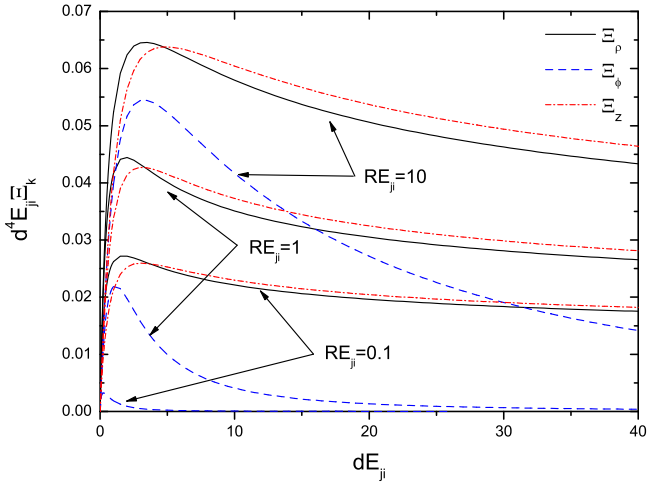,
width= 10cm}}
\end{center}
\vspace*{-10mm} \caption{The contributions (\ref{FinalC1})--(\ref{FinalC3}) to the energy
  shift (\ref{generalShift}) due to the $\rho$, $\phi$, and $z$ components
  of the dipole for various radii $R$ of the wire.}
\label{fig:combined}
\end{figure}

In Fig.~\ref{fig:combined} we show how these contributions look when we
choose the wavelengths $1/E_{ji}$ of a typical internal transition as a
lengthscale and plot the contributions to the energy shift for various wire
radii $R$. The larger the value of $R$ the more terms are required in the
numerical series.

\section{ENERGY SHIFT NEAR A PERFECTLY REFLECTING SEMI-INFINITE HALFPLANE}
\begin{figure}
\begin{center}
\centerline{\epsfig{figure=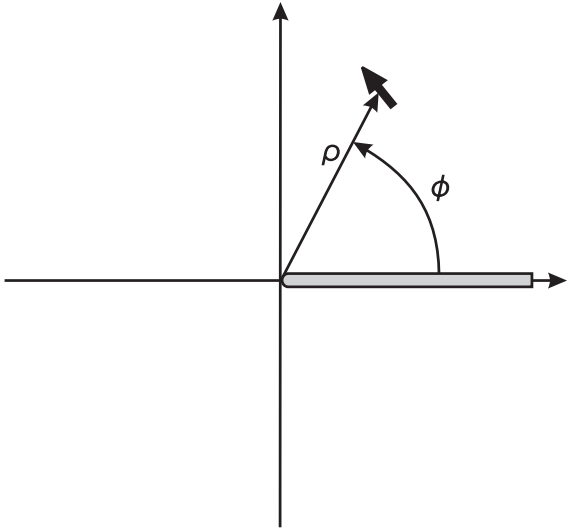,
width= 7.5cm, height= 7.0 cm}}
\end{center}
\vspace*{-10mm} \caption{An atomic dipole in the vicinity of a perfectly
  reflecting semi-infinite halfplane. The normal modes
  $\mathbf{F}_\lambda^{(\sigma)}(\bx)$ in this geometry are given by
  Eqs. (\ref{HalfPlaneTE}) and (\ref{HalfPlaneTM}).}
\label{fig:3}
\end{figure}
Next we wish to calculate the energy shift of an atom in the vicinity of a
perfectly reflecting halfplane, as illustrated by Fig. \ref{fig:3}. 

The procedure of obtaining the normal modes of the vector potential is
analogous to that described in Section \ref{sec3}. The scalar solution of
the Helmholtz equation (\ref{helm}) in the cylindrical coordinates that is
best suited to applying boundary conditions on the surface of the halfplane
is given by
\begin{equation}
\Phi(\bx)=\bigg[\frac{\alpha}{\sqrt\pi} \sin 
\left(\frac{m}{2}\phi\right)+\frac{\beta}{\sqrt\pi} 
\cos\left(\frac{m}{2}\phi\right) \bigg]J_{m/2}(k\rho)
\;\frac{e^{{\rm i}\kappa z}}{\sqrt{2\pi}}\;,
\nonumber
\end{equation}
where $J_{m/2}(k\rho)$, with $m=0,1,2,\ldots$, are the regular solutions of
Bessel's equation, and the separation constants satisfy
$\omega^2=k^2+\kappa^2$. We must have $m\geq 0$, as otherwise the solutions
are not linearly independent. Note that half-integer indices arise because
the angle $\phi$ is restricted to the interval $[0,2\pi]$, so that the usual
argument of single-valuedness of $e^{{\rm i}m\phi}$ cannot be evoked.

In order to obtain two linearly independent vector solutions we again apply
Eqs. ~(\ref{sol1}) and (\ref{sol2}), and impose the boundary conditions for
a perfectly reflecting halfplane, $E_\rho=0=E_z$ and $B_\phi=0$ for $\phi=0$
and $\phi=2\pi$. In this way we find for the mode functions
\begin{eqnarray}
\mathbf{F}_\lambda^{(1)}(\br) &\!\!=&\!\! -\frac{1}{\sqrt{2}\pi}
\bigg[\frac{m}{2k\rho}\sin\left(\frac{m}{2}\phi\right)J_{m/2}(k\rho)\;
\hat{\mathbf{e}}_\rho\nonumber\\
 &\!\!+&\!\! \cos\left(\frac{m}{2}\phi\right)J'_{m/2}(k\rho)\;
\hat{\mathbf{e}}_\phi\bigg]e^{{\rm i}\kappa z},
\label{HalfPlaneTE} \\
\mathbf{F}_\lambda^{(2)}(\br) &\!\!=&\!\!  \frac{1}{\sqrt{2}\pi}
\bigg[\frac{i\kappa}{\omega}\sin\left(\frac{m}{2}\phi\right)J'_{m/2}(k\rho)\;
\hat{\mathbf{e}}_\rho\nonumber\\
 &\!\!+&\!\! \frac{i\kappa m}{2k\rho\omega}\cos\left(\frac{m}{2}\phi\right)
J_{m/2}(k\rho)\;\hat{\mathbf{e}}_\phi\nonumber\\
 &\!\!+&\!\! \frac{k}{\omega}\sin\left(\frac{m}{2}\phi\right)J_{m/2}(k\rho)\;
\hat{\mathbf{e}}_z\bigg]e^{{\rm
 i}\kappa z},
\label{HalfPlaneTM}
\end{eqnarray}
where the composite index stands for $\lambda=\{k,m,\kappa\}$.  For $m>0$
these mode functions satisfy the normalization condition (\ref{vectornorm}),
but the first polarization has an additional mode with $m=0$ for which
Eq.~(\ref{HalfPlaneTE}) must be multiplied by an additional factor
$1/\sqrt{2}$ for it to be normalized correctly according to
(\ref{vectornorm}),
\begin{equation}
\mathbf{F}_{m=0}^{(1)}(\br) = -\frac{1}{2\pi}
J'_{0}(k\rho)\;\hat{\mathbf{e}}_\phi \;e^{{\rm i}\kappa z}\;.
\label{Fm0}
\end{equation}

Substituting the mode functions (\ref{HalfPlaneTE})-(\ref{Fm0}) into
Eq.~(\ref{dW}) and renormalizing the energy shift by subtracting the
free-space contribution in the same way as this was done in
Eqs.~(\ref{FreeKsiRho})-(\ref{FreeKsiZ}), we obtain an energy shift of the
form (\ref{generalShift}) with
\begin{widetext}
\begin{eqnarray}
\Xi _\rho &\!\!=&\!\! \frac2\pi \int_0^\infty \rd k\;k\int_{0}^{\infty}\rd
\kappa\;\frac{\omega}{E_{ji}+\omega}
\left\{ 
 \left(\frac{1}{k\rho}\right)^2\sum_{m=1}^\infty \left[ 
\left(\frac{m}{2}\right)^2\sin^2\left(\frac{m}{2}\phi\right)J^2_{m/2}(k\rho)
-m^2J^2_m(k\rho)\right]\right.
\nonumber \\ 
 &\!\!+&\!\!\left.\left(\frac{\kappa}{\omega}\right)^2\sum_{m=0}^\infty\!\:'
\left[ \sin^2\left(\frac{m}{2}\phi\right)J'^2_{m/2}(k\rho)
-J'^2_{m}(k\rho) \right]
\right\},
\label{KsiRho2}
\\
\Xi _\phi &\!\!=&\!\! \frac2\pi \int_0^\infty \rd k\;k\int_{0}^{\infty}\rd
\kappa\;\frac{\omega}{E_{ji}+\omega}
\left\{ 
 \sum_{m=0}^\infty\!\:' \left[ \cos^2\left(\frac{m}{2}\phi\right)
J'^2_{m/2}(k\rho)
-J'^2_m(k\rho)\right]\right.
\nonumber \\ 
 &\!\!+&\!\!\left. \left(\frac{\kappa}{k\rho\omega}\right)^2\sum_{m=1}^\infty 
\left[ \left(\frac{m}{2}\right)^2\cos^2\left(\frac{m}{2}\phi\right)
J^2_{m/2}(k\rho)
-m^2J^2_m(k\rho) \right]
\right\},
\label{KsiPhi2}
\\
\Xi _z &\!\!=&\!\! \frac2\pi \int_0^\infty \rd k\;k\int_{0}^{\infty}\rd
\kappa\;\frac{\omega}{E_{ji}+\omega}
\left\{ 
 \left(\frac{k}{\omega}\right)^2\sum_{m=0}^\infty \!\:' \left[ \sin^2\left(
\frac{m}{2}\phi\right)J^2_{m/2}(k\rho)
-J^2_m(k\rho)\right]\right\},
\label{KsiZ2}
\end{eqnarray}
\end{widetext}
where the primes on the sums indicate that the $m=0$ terms are weighted by
an additional factor of 1/2. In order to simplify these expressions, the sums
over the Bessel functions need to be evaluated. Recently, similar summations
have been carried out \cite{Farina2,Farina1}, but the results obtained
do not include our particular case of sums involving Bessel functions of
the half-integer order.

We proceed along the following lines. First, we split each sum into two,
one over Bessel functions of integer orders, and the other over half-integer
orders. For the first we can apply the standard summation formula 
\cite[9.1.79]{AS}
\begin{equation}
\sum_{m=0}^{\infty}\!{}'\; \cos 2m\phi\;J_m^2(z)= 
\frac{1}{2} J_0(2 z \sin\phi)\;,
\label{intsum}
\end{equation}
and we choose to represent the right-hand side in terms of an integral
\cite[9.1.24]{AS}
\begin{equation}
\frac{1}{2} J_0(2 z \sin\phi) = \frac{1}{\pi} \int_1^\infty \rd t 
\;\frac{\sin(2zt\sin\phi)}{\sqrt{t^2-1}}\;.
\label{intrep}
\end{equation}
For the half-integer sum we use a summation formula of
\cite[5.7.17.(11.)]{Prudnikov}, which in our case gives
\begin{equation}
\sum_{m=0}^{\infty}\cos (2m+1)\phi\;J_{m+\frac12}^2(z)= \frac{1}{\pi} 
\int_1^{1/\sin\phi} \!\!\rd t \;\frac{\sin(2zt\sin\phi)}{\sqrt{t^2-1}}\;.
\label{hintsum}
\end{equation}
We note that, if we use the integral representation (\ref{intrep}), the sums
over integer and over half-integer Bessel functions are very similar; the
only difference is the upper limit of the $t$ integral in (\ref{intrep}) and
(\ref{hintsum}). As these $t$ integrals and their derivatives will arise
repeatedly, we define the following auxiliary functions:
\begin{eqnarray}
&&F(z,\phi)\equiv 
\int_1^{1/\sin\phi} \!\!\rd t \;\frac{\sin(2zt\sin\phi)}{\sqrt{t^2-1}}\;,
\label{F}\\
&&G(z,\phi)\equiv 
\int_1^\infty \rd t \;\frac{\sin(2zt\sin\phi)}{\sqrt{t^2-1}}\;.
\label{G}
\end{eqnarray}

Further we note that the $\kappa$ integrals in
Eqs. (\ref{KsiRho2})-(\ref{KsiZ2}) suffer from the same convergence problems
as already discussed in Section \ref{sec3}. We avoid these by introducing
polar coordinates with  ${k=\omega\sin\alpha}$ and
${\kappa=\omega\cos\alpha}$. At the same time we parametrize the denominator
arising from perturbation theory by
\begin{eqnarray}
\frac{1}{E_{ji}+\omega}=\int_0^\infty \rd x\;e^{-(E_{ji}+\omega)x}\;,
\label{para}
\end{eqnarray}
with $E_{ji}+\omega=E_{ji}+\sqrt{k^2+\kappa^2}\geq0$.
Then Eqs. (\ref{KsiRho2})-(\ref{KsiZ2}) become
\begin{eqnarray}
\Xi _\rho &\!\!=&\!\! \frac2\pi\int_0^\infty \rd x\;e^{-E_{ji}x}
\int_0^\infty \rd\omega\;\omega^3e^{-\omega x}
\int_0^{\pi/2}\rd\alpha\sin\alpha
\nonumber \\
&&\hspace*{5mm}\times\left\{ \sigma_1(\omega\rho\sin\alpha) +  
\sigma_3(\omega\rho\sin\alpha)\cos^2\alpha
\right\},
\label{aux1}
\end{eqnarray}
\begin{eqnarray}
\Xi _\phi &\!\!=&\!\! \frac2\pi\int_0^\infty \rd x\;e^{-E_{ji}x}
\int_0^\infty \rd \omega\;\omega^3e^{-\omega x}
\int_0^{\pi/2}\rd\alpha\sin\alpha
\nonumber \\
&&\hspace*{5mm}\times\left\{ \sigma_2(\omega\rho\sin\alpha)\cos^2\alpha
+\sigma_4(\omega\rho\sin\alpha)
\right\},\label{aux2}
\end{eqnarray}
\begin{eqnarray}
\Xi _z &\!\!=&\!\! \frac2\pi\int_0^\infty \rd x\;e^{-E_{ji}x}
\int_0^\infty \rd \omega\;\omega^3e^{-\omega x}
\int_0^{\pi/2}\rd\alpha\sin\alpha
\nonumber \\
&&\hspace*{5mm}\times\left\{ \sigma_5(\omega\rho\sin\alpha)\sin^2\alpha
\right\}.\label{aux3}
\end{eqnarray}
The sums $\sigma_i(z)$ appearing in these expressions can be calculated by
using Eqs.~(\ref{intsum})--(\ref{G}) and standard derivative formulae for
Bessel functions \cite[9.1.27]{AS}; we obtain in terms of (\ref{F}) and
(\ref{G}):
\begin{widetext}
\begin{eqnarray}
\left[\begin{array}{c}\sigma_1(z)\\\sigma_2(z)\end{array}\right]
&\!\!=&\!\! \frac{1}{z^2} \sum_{m=1}^\infty 
\left\{\left(\frac{m}{2}\right)^2\left[
\begin{array}{c}\sin^2\left(m\phi/2\right)\\
\cos^2\left(m\phi/2\right)\end{array}\right]
J^2_{m/2}(z)-m^2J^2_{m}(z)\right\}\nonumber\\
&\!\!=&\!\!\frac{1}{8\pi z^2} \left[ \pm 
\frac{\partial^2G(z,\phi)}{\partial\phi^2}+
\frac{\partial^2G(z,\phi)}{\partial\phi^2}\bigg|_{\phi=0}\pm 
\frac{\partial^2F(z,\phi)}{\partial\phi^2}
-\frac{\partial^2F(z,\phi)}{\partial\phi^2}\bigg|_{\phi=0}\right],
\label{sigma1}\\
\left[\begin{array}{c}\sigma_3(z)\\\sigma_4(z)\end{array}\right]
&\!\!=&\!\! \sum_{m=0}^\infty\!\!\;'
\left\{\left[\begin{array}{c}\sin^2\left(m\phi/2\right)\\
\cos^2\left(m\phi/2\right)\end{array}\right]
J'^2_{m/2}(z)-J'^2_{m}(z)\right\}\nonumber\\
&\!\!=&\!\!- 
\left[\begin{array}{c}\sigma_1(z)\\\sigma_2(z)\end{array}\right]
+\frac{1}{2\pi}\big[F(z,0)-G(z,0)\big]\mp\frac{\cos 2\phi}{2\pi}
\big[F(z,\phi)+G(z,\phi)\big]+\frac{\cos 2z}{2\pi z}(1\mp\cos\phi)\;,
\label{sigma2}\\
\sigma_5(z)&\!\!=&\!\! \sum_{m=0}^\infty\!\!\;'
\left\{ \sin^2\left(m\phi/2\right)J^2_{m/2}(z)-J^2_{m}(z)
\right\}\nonumber\\
&\!\!=&\!\! \frac{1}{2\pi}\big[F(z,0)-F(z,\phi)-G(z,\phi)-G(z,0) \big]\;.
\label{sigma3}
\end{eqnarray}
\end{widetext}
We now carry out the various integrations in the following order. First we
evaluate the $\alpha$ integrals, which all give Bessel functions $J_1$ or
$J_0$ \cite[3.715(10),(14)]{GR}. Next we carry out the integrations over 
$\omega$, which involve integrals of the type \cite[6.611(1.)]{GR} 
\begin{equation}
\int_0^\infty \rd z\; e^{-az}\,J_\nu(bz)=
\frac{b^{-\nu}\left(\sqrt{a^2+b^2}-a\right)^\nu}{\sqrt{a^2+b^2}}\;.
\nonumber
\end{equation}  
Finally, we calculate the $t$ integrals that came in through the auxiliary
functions $F$ and $G$, Eqs. (\ref{F}) and (\ref{G}). These are all
elementary. At the very end we calculate the $\phi$ derivatives of
Eq.~(\ref{sigma1}) and take the limit $\phi\rightarrow 0$ in the appropriate
terms. The end results then still contain the parameter integral
(\ref{para}) over $x$, which we now scale by substituting $x=2\rho\eta$.
Then the final results read
\begin{widetext}
\begin{eqnarray}
\Xi _\rho&\!\!=&\!\!\frac{1}{16\pi\rho^3}\int_0^\infty \rd\eta\; 
e^{-2\rho E_{ji}\eta}\left\{\frac{3\eta^4+6\eta^2+4}{\eta^4(1+\eta^2)^{3/2}}
-\frac{4}{\eta^4}+\frac{4}{(\eta^2+\sin^2\phi)^3}\left[(2\eta^2+1)\sin^2\phi
-\eta^2\right]\right.\nonumber\\
&\!\!+&\!\!\left.\frac{\cos\phi}{(1+\eta^2)^{3/2}(\eta^2+\sin^2\phi)^3}
\left[(2+\eta^2)\sin^4\phi+2\sin^2\phi(3\eta^4+6\eta^2+2)-\eta^2
(3\eta^4+6\eta^2+4)\right]\right\}, 
\label{Final1}\\ \nonumber\\
\Xi _\phi&\!\!=&\!\!\frac{1}{16\pi\rho^3}\int_0^\infty \rd\eta\; 
e^{-2\rho E_{ji}\eta}\left\{\frac{3\eta^6+6\eta^4+10\eta^2+4}{\eta^4
(1+\eta^2)^{5/2}}-\frac{4}{\eta^4}+\frac{4}{(\eta^2+\sin^2\phi)^3}
\left[(1-2\eta^2)\sin^2\phi+\eta^2\right]
\right.
 \nonumber\\
&\!\!+&\!\!\left.\frac{\cos\phi}{(1+\eta^2)^{5/2}(\eta^2+\sin^2\phi)^3}
\left[(2-2\eta^2-\eta^4)\sin^4\phi+2\sin^2\phi(2+2\eta^2-6\eta^4-3\eta^6)
+\eta^2(3\eta^6+6\eta^4+10\eta^2+4)\right]
\right\},\nonumber\\
\label{Final2}\\
\Xi _z&\!\!=&\!\!\frac{1}{16\pi\rho^3}\int_0^\infty \rd\eta\; 
e^{-2\rho E_{ji}\eta} \left\{\frac{9\eta^4+10\eta^2+4}{\eta^4(1+\eta^2)^{5/2}}
-\frac{4}{\eta^4}+4\frac{\sin^2\phi-\eta^2}{(\eta^2+\sin^2\phi)^3}
\right.
\nonumber \\ 
&\!\!-&\!\!\left.\frac{\cos\phi}{(1+\eta^2)^{5/2}(\eta^2+\sin^2\phi)^3}
\left[(\eta^2-2)\sin^4\phi+2(\eta^4-4\eta^2-2)\sin^2\phi
+\eta^2(9\eta^4+10\eta^2+4)\right]\right\}.
\label{Final3}
\end{eqnarray}
\end{widetext}
Inserted into Eq.~(\ref{generalShift}) the
Eqs.~(\ref{Final1})--(\ref{Final3}) give the final result for the energy
shift of an atom near a perfectly reflecting halfplane. Some of the
integrations over the auxiliary variable $\eta$ could in principle be
carried out, but those would yield complicated hypergeometric
functions. Thus it is preferable to have the result in the form of an
integral over elementary functions. It converges quickly and can therefore
be very easily evaluated numerically by using standard software packages. In
addition, we shall go on to determine asymptotic expressions in the
non-retarded and retarded regimes.

\subsection{Asymptotic regimes.}
\subsubsection{Plane-mirror limit}\label{planelimit}
In the limit of the polar angle $\phi$ being very small, the atom is very
close to the halfplane but far away from the edge, so that the energy shift
should be the same as for an atom in front of a plane, infinitely extended
mirror. The component of the atomic dipole that is normal to the surface
should then give the contribution listed in Eq.~(\ref{CPperp}) to the shift,
and the parallel components should contribute that shown in Eq.~(\ref{CPpara}).
As the distance $d$ of the atom from the halfplane is $\rho\sin\phi$, we
take Eqs.~(\ref{Final1})--(\ref{Final3}) and scale
$\eta\rightarrow\eta\sin\phi$, so as to get an exponential with the same
argument as in Eqs.~(\ref{CPperp}) and (\ref{CPpara}). If we subsequently
take the limit $\phi\rightarrow 0$, we recover Eqs.~(\ref{CPperp}) and
(\ref{CPpara}), as expected. Note, however, that the geometry is different
from the cylindrical case: the $\phi$ component of the atomic dipole is now
normal to the surface and its contribution $\Xi_\phi$ to the energy shift is
given by (\ref{CPperp}), and the $\rho$ and $z$ components are parallel so
that $\Xi_\rho$ and $\Xi_z$ are given by (\ref{CPpara}).

\subsubsection{Non-retarded regime}
If $\rho E_{ji}\ll 1$ then the atom is very close to the halfplane, compared
to the wavelength of a typical internal transition. This means that the
interaction of the atom and the surface is instantaneous, as the atom
evolves on a much longer timescale. In this case field quantization is not
necessary, and only Coulomb interactions between the atom and the halfplane
need to be considered, as was done in Ref.~\cite{nonret}, where we derived
\begin{eqnarray}
\Xi_\rho&\!\!=&\!\!\frac{5}{48\pi\rho^3}+\frac{\cos\phi}{16\pi\rho^3\sin^2\phi}
+\frac{(\pi-\phi)(1+\sin^2\phi)}{16\pi\rho^3\sin^3\phi}\nonumber\\
\Xi_\phi&\!\!=&\!\!-\frac{1}{48\pi\rho^3}
+\frac{\cos\phi}{8\pi\rho^3\sin^2\phi}
+\frac{(\pi-\phi)(1+\cos^2\phi)}{16\pi\rho^3\sin^3\phi}\nonumber\\
\Xi_z&\!\!=&\!\!\frac{1}{24\pi\rho^3}+\frac{\cos\phi}{16\pi\rho^3\sin^2\phi}
+\frac{\pi-\phi}{16\pi\rho^3\sin^3\phi}\nonumber\;.
\end{eqnarray}
Taking the limit $E_{ji}\rightarrow 0$ in
Eqs.~(\ref{Final1})--(\ref{Final3}) we recover these results, which is an
important consistency check on our present calculation.
 
\subsubsection{Retarded regime}
In the opposite limit of the atom being far away from the halfplane, we need to
distinguish whether the atom is located beyond the edge of the halfplane or
not. If it is, i.e. for $\pi/2<\phi<\pi$ the distance of the atom to the
halfplane is its distance to the edge, namely $\rho$, so that the condition
for the interaction to be fully retarded is $\rho E_{ji}\gg 1$. If, on the
other hand, $0<\phi<\pi/2$ then the distance to the halfplane is
$\rho\sin\phi$, and consequently the criterion for full retardation is 
$\rho\sin\phi E_{ji}\gg 1$, cf.~Fig.~\ref{fig:3}.

Taking the limit $E_{ji}\rightarrow\infty$ in the integrals
(\ref{Final1})-(\ref{Final3}) is straightforward, since, according to
Watson's lemma \cite{Bender}, the integral is then dominated by
contributions from the vicinity of $\eta=0^+$, so that one just needs to
factor out the exponential and expand the rest of the integrand {in the
curly brackets} in a Taylor series about this point. The leading terms of
these Taylor expansions turn out to be constants with respect to $\eta$ in
each case. Thus in the retarded limit we obtain
\begin{eqnarray}
\Xi _\rho &\!\!=&\!\!\frac{1}{64\pi\rho^4 E_{ji}}\left[3
+\frac{1}{\sin^4(\phi/2)}+\frac{2}{\sin^2(\phi/2)}
\right]\label{ret1}\\
\Xi _\phi &\!\!=&\!\!\frac{1}{64\pi\rho^4 E_{ji}}\left[ -3
+\frac{1}{\sin^4(\phi/2)}+\frac{2}{\sin^2(\phi/2)}
\right]\label{ret2}\\
\Xi _z &\!\!=&\!\! \frac{1}{64\pi\rho^4 E_{ji}}\left[3
+\frac{1}{\sin^4(\phi/2)}+\frac{2}{\sin^2(\phi/2)}
\right],\label{ret3}
\end{eqnarray}
which, for the case of isotropic polarizability, is in agreement with the
result of Ref.~\cite{Brevik}. In the light of our comments above, we
emphasize again that these results are only valid when the distance of the
atom from the halfplane exceeds several wavelengths $\lambda_{ji}$. This
means that for small angles $\phi$ one needs to revert to the plane-mirror
limit discussed in Section \ref{planelimit} above, because in the region
$0<\phi<\pi/2$ Eqs.~(\ref{ret1})--(\ref{ret3}) apply only if $\sin\phi\gg
\lambda_{ji}/\rho$.  However, taking the limit $\rho\rightarrow\infty$
together with $\phi\rightarrow0$ while keeping $\rho\sin\phi=d$ fixed is
legitimate, and reproduces the well-known Casimir-Polder result
\cite{casimir} for the retarded interaction between an atom and a plane
mirror, Eq. (\ref{PlaneCP}).

Taking the limit $\phi\rightarrow\pi$ in Eqs.~(\ref{ret1})--(\ref{ret3})
shows that for an atomic dipole that is polarized azimuthally the
interaction vanishes when the atom is located exactly above the edge of the
halfplane. This conclusion actually holds not just in the retarded regime,
but generally for any distance, as Eq. (\ref{Final2}) also vanishes in the
limit $\phi\rightarrow\pi$. Purely from symmetry one would expect there to
be no azimuthal component to the Casimir-Polder force directly above the
edge, but the fact that there is no radially directed force either is
surprising. 
\begin{figure}[t]
\vspace*{3mm}
\begin{center}
\centerline{\epsfig{figure=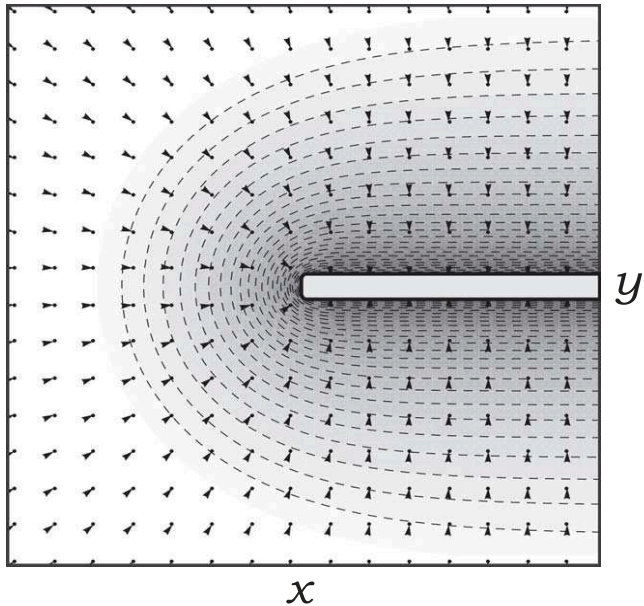,
width= 8.5 cm, height= 8.0 cm}}
\end{center}
\vspace*{-5mm} \caption{Direction of the retarded Casimir-Polder force
acting on the atom with isotropic polarizability. Note from
Eqn. (\ref{Final2}) that an atom that is polarized azimuthally does not
experience any force when it is located exactly above the edge of the 
halfplane.}
\label{fig:4}
\end{figure}

Since we have worked in the cylindrical coordinates, the direction of the
unit vectors $\hat{\mathbf{e}}_\rho$ and $\hat{\mathbf{e}}_\phi$ depends on
the position coordinates $\rho$ and $\phi$. In this context it is curious
that, in the retarded limit, all three components of the atomic dipole
contribute to the energy shift with exactly the same angular dependence.

\section{summary}
We have calculated the energy shift in a neutral atom caused by the presence
at arbitrary distance of perfectly reflecting microstructures of two
different geometries. For an atom at a distance $d=\rho-R$ from the
perfectly reflecting cylindrical wire of radius $R$ we have found an exact
expression for the interaction energy, Eq.~(\ref{generalShift}) with
Eqs. (\ref{FinalC1})-(\ref{FinalC3}). As these integrals and sums are in
general quite complicated, we have analysed various important limiting
cases. The limit of the distance $d$ being small on the scale of the
wavelength $\lambda_{ji}$ of a typical atomic transition requires only
electrostatic forces to be considered, which was done in detail in
Ref.~\cite{nonret}. The case of purely retarded interactions, which occur
when the distance $d$ is much larger than $\lambda_{ji}$, has been analysed
in Sections \ref{case3}--6. For a small wire radius the three contributions
to the energy shift are well approximated by Eqs.~(\ref{A2})--(\ref{C2}), and
for a large wire radius by Eqs.~(\ref{A})--(\ref{C}).

In the case of an atom close to a perfectly reflecting halfplane the exact
analytic analysis can be pushed a little bit further than in the cylindrical
case. We have managed to find an exact formula for the energy shift in terms
of a simple, rapidly converging integral over elementary functions,
Eqs. (\ref{Final1})-(\ref{Final3}), so that they are very easy to study
numerically. Nevertheless, we have also derived asymptotic formulae, which
agree with previous calculations.

The totality of our results can be used to reliably estimate the energy
shift in an atom close to a variety of common microstructures that consist
of a ledge and possibly an electroplated top layer of higher
reflectivity. We have determined the energy shifts for the
complete range of distances, which is very important for practical
applications as in many modern experiments the distance of the atom is
neither much larger nor much smaller than the typical wavelength of an atomic
transition, but commensurate.

\begin{acknowledgments}
It is a pleasure to thank Gabriel Barton for discussions. 
We would like to acknowledge financial support from the UK Engineering and
Physical Sciences Research Council.
\end{acknowledgments}

\end{document}